\documentclass[twocolumn,showpacs,prb,aps,amssymb]{revtex4}

\usepackage{graphicx}
\usepackage{bm}

\begin{document}

\newcommand{\mysw}[1]{{\scriptscriptstyle #1}}

\draft
\title{Ground-state properties of nanographite systems
with zigzag-shaped edges}

\author{Toshiya Hikihara and Xiao Hu}
\affiliation{Computational Materials Science Center, 
National Institute for Materials Science, Tsukuba,
Ibaragi 305-0047, Japan}
\author{Hsiu-Hau Lin$^{1,2}$ and Chung-Yu Mou$^{1}$}
\affiliation{$^1$ Department of Physics, National Tsing-Hua University, 
Hsinchu 300, Taiwan\\
$^2$ Physics Division, National Center for Theoretical Science, 
Hsinchu 300, Taiwan}
\date{\today}

\begin{abstract}
A $\pi$-electron network in nanographite systems with zigzag edges 
exhibits strongly localized edge states, 
which are expected to have peculiar properties.
We study effects of electron-electron interactions 
on ground-state properties of the systems 
by means of the weak-coupling renormalization group and 
the density-matrix renormalization-group method.
It is shown that the ground state is a spin-singlet Mott insulator 
with finite charge and spin gaps.
We also find that the edge states are robust against 
the electronic correlations, resulting in 
edge effective spins which can flip almost freely.
The schematic picture for the low-energy physics 
of the systems is discussed.
\end{abstract}

\maketitle

\section{Introduction}

Since the discoveries of fullerenes\cite{C60} 
and carbon nanotubes (CNTs),\cite{Iij} 
the nature of ``nanographites", 
graphite-based materials with nano-meter sizes, 
has attracted much attention from both fundamental science and applications.
One of the most striking features of the materials 
is a wide variety of properties they show depending on their geometrical 
structure, i.e., size, shape, surface condition and so on.
For example, CNTs can be either metallic or semiconducting 
depending on the wrapping superlattice vector.\cite{review}
Such a diversity of possible functions realized in nanographites  
makes them promising candidates for nano-scale devices.

Recently, it has been pointed out that a $\pi$-electron system 
in nanographites with zigzag-shaped edges exhibits 
peculiar electronic states strongly localized around the edges, 
which are termed ``edge states".\cite{MFT1,edge1,edge2}
Such a state does not appear around armchair-shaped edges.
The nature of the edge states has been studied 
in the graphene sheet of nano-meter width, 
named the ``nanographite ribbon" (NGR).
Applying a tight-binding model for the NGR with zigzag edges, 
Fujita {\it et al.} have shown that the system 
has the electronic states localized on the edges, which penetrate 
from the edges into the bulk decaying exponentially.\cite{MFT1,edge1}
The edge states are believed to be responsible for 
a paramagnetic behavior observed in activated carbon fibers,
which are taken to be an assembly of nanographite particles.\cite{ACF1,ACF2} 

Since the edge states are characterized by almost flat dispersions 
in a certain range of the momentum space, 
which result in a sharp peak in the density of state at the Fermi energy,
possible instabilities to various perturbations are important and 
have been investigated to date.
Concerning the Peierls instability due to electron-phonon interactions,
it has been shown that no bond alternation occurs in the zigzag NGRs 
with realistic interaction strength.\cite{el-ph}
On the other hand, effects of electron-electron interactions on 
the edge states have been investigated 
by a mean-field approximation.\cite{MFT1,MFT2}
The studies have shown that 
an infinitesimal on-site interaction causes 
a spontaneous spin polarization at the zigzag edges.
Similar results have been obtained for the zigzag CNTs 
by the density-functional theory.\cite{DFT1,DFT2}
However, it is also known that 
the one-body approximation adopted in these approaches 
is not appropriate for one-dimensional (1D) quantum systems 
such as NGRs and CNTs.
In fact, Lieb's theorem\cite{Lieb} prohibits 
the spontaneous spin polarization for the Hubbard model
in the nanographite systems with zigzag edges.
More detailed analyses with controlled approximations 
are therefore desirable for clarifying the low-energy properties 
of the nanographites.

In this paper, we study effects of electron-electron interactions 
on the nanographite systems with zigzag edges 
using two powerful techniques: 
the weak-coupling renormalization group (RG) and 
the density-matrix renormalization group (DMRG) method.\cite{White1}
These approaches allow us to treat the strong quantum fluctuations in 
a controlled way for the 1D systems such as NGRs 
and CNTs. With the inclusion of electronic correlations, we show 
that the ground state of the zigzag NGR is a gapped spin-singlet and
the artificially broken symmetry of spin rotations in the mean-field type 
analysis is restored by quantum fluctuations. 
It is rather interesting that the edge states survive in the presence of 
electron-electron interactions and play a crucial role 
in the low-energy regime.

The rest of this paper is organized as follows.
The model is presented in the next section.
In Sec. III, we show results of the weak-coupling analysis.
We first discuss properties of the tight-binding model of the NGR system.
Analytical eigen-wavefunctions and the corresponding energy spectrum 
are shown.
And then, by keeping the low-lying edge states only and taking 
the electron-electron couplings into account, 
we derive the low-energy effective Hamiltonian in the continuum limit.
It is shown that all coupling terms included in the Hamiltonian are relevant 
because of a large dynamical exponent of the dispersion of the edge states.
The enhanced density of states at the Fermi energy generates 
finite charge and spin gaps, and consequently, 
the system exhibits a spin-singlet ground state 
without spontaneous symmetry breaking.
The DMRG results are presented in Sec. IV.
We find that in the presence of the Hubbard interaction 
the ground state is a spin singlet with finite charge and spin gaps, 
and no spontaneous spin polarization appears.
Besides, electrons around the zigzag edges correlate 
ferromagnetically to compose effective spins which are easily polarized 
by applying an external field, 
while electrons in the bulk form a spin-singlet state 
which are hardly magnetized.
The effective spins are localized around the zigzag edges 
even for a rather strong coupling, 
showing that the localization character of the edge states 
persists robustly against the electron-electron interactions.
The schematic picture for the low-energy physics 
of the systems is discussed.
Finally, our results are summarized in Sec. V.

\section{Model Hamiltonian}
\begin{figure}
\includegraphics[width=60mm]{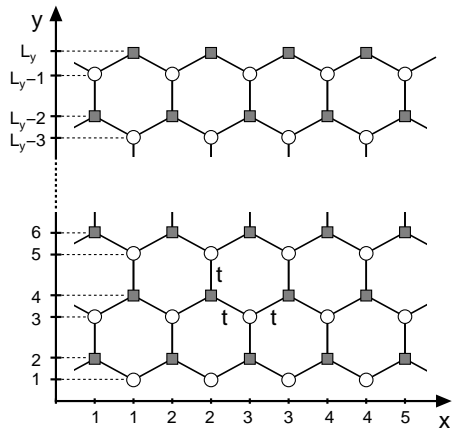}
\caption{
Honeycomb lattice of carbon atoms in 
the zigzag NGRs and CNTs. 
The open circles and gray squares represent 
carbon sites of sublattice A and B, respectively.} 
\label{fig:lattice}
\end{figure}

In this paper, we consider a $\pi$-electron system at half-filling 
on a honeycomb lattice with zigzag edges as shown in Fig.~\ref{fig:lattice}.
The effect of electron-electron couplings is incorporated 
by introducing the on-site Hubbard interaction $U$.
The Hamiltonian is 
\begin{eqnarray}
{\cal H} &=& {\cal H}_0 + {\cal H}_U,  \label{eq:Ham} \\
{\cal H}_0 &=& 
- \sum_{{\bm r},{\bm r'}} \sum_{\alpha = \uparrow,\downarrow}
   t({\bm r}, {\bm r'}) 
   \left[ c^\dagger_\alpha ({\bm r}) c_\alpha ({\bm r'}) + h.c. \right],
\label{eq:Ham0} \\
{\cal H}_U &=& 
 U \sum_{\bm r}
 n_\uparrow ({\bm r}) n_\downarrow ({\bm r}),
\label{eq:HamU}
\end{eqnarray}
where $c_\alpha ({\bm r})$ ($c^\dagger_\alpha ({\bm r})$) is the fermion 
annihilation (creation) operator at the site ${\bm r} = (x,y)$ and
$n_\alpha ({\bm r}) \equiv c^\dagger_\alpha ({\bm r}) c_\alpha ({\bm r})$.
The definition of the site index is shown in Fig.~\ref{fig:lattice}.
The system size in the $x$- and $y$-directions are denoted by 
$L_x$ and $L_y$, respectively.
Within tight-binding approximation, the hopping amplitude 
$t({\bm r}, {\bm r'})$ is assumed to be 
$t$ between nearest-neighbor sites and otherwise $0$.
In the $x$-direction, the open boundary condition with $L_x \to \infty$ 
is imposed for the zigzag NGR while the periodic boundary condition 
is imposed for the zigzag CNT.
In the $y$-direction, the open boundary condition is imposed on 
the zigzag edges. We set $L_y$ to be even so that the system always has 
the reflection symmetry in the $y$-direction.
All carbon sites at the edges are assumed to be terminated by hydrogen atoms. 

We note that the system is bipartite and the number of sites 
in each sublattice is equal, $N_A = N_B$. At the natural filling, with 
one $\pi$-electron per site on average, the system is particle-hole 
symmetric. Thus, one can apply Lieb's theorem\cite{Lieb} to the system, 
which prohibits spontaneous spin polarizations in the ground state.

\section{Renormalization group}
\subsection{Band structure}
To study the system in the weak-coupling limit $U \ll t$, 
it is natural to consider the case of $U = 0$ and analyze 
the band structure first.
We thus begin with diagonalizing the hopping Hamiltonian ${\cal H}_0$.
Since the Hamiltonian is translational invariant along the $x$-direction, 
we can perform the partial Fourier transformation,\cite{Lin}
\begin{equation}
c_\alpha (k_x; y) 
= \frac{1}{\sqrt{L_x}} \sum_x \exp[-ik_x (x + \delta)] c_\alpha (x, y)
\label{eq:FT0}
\end{equation}
where $\delta = 1/2$ for $y = 0, 1$ (mod 4) 
and $\delta = 0$ for $y = 2, 3$ (mod 4).
By the transformation, 
the hopping Hamiltonian ${\cal H}_0$ is mapped into 
bond-alternating chains decoupled from each other,
\begin{eqnarray}
\tilde{\cal H}_0 &=& \sum_{k_x} \tilde{\cal H}_0(k_x), 
\nonumber \\
\tilde{\cal H}_0(k_x) &=&
- \sum_{y=1}^{L_y-1} \sum_\alpha
\tilde{t}(y) [c^\dagger_\alpha (k_x; y+1) c_\alpha (k_x; y) + {\rm h.c.} ],
\nonumber \\
\end{eqnarray}
where $\tilde{t}(y) = t_1 \equiv 2 t \cos(k_x/2)$ if $y$ is odd 
while $\tilde{t}(y) = t_2 \equiv t$ if $y$ is even.
It is important to note that the effective hopping $t_1$ depends on 
the momentum $k_x$, and consequently, 
the motions in the $x$- and $y$-directions are entangled.

Each of the bond-alternating chains is composed of 
two sublattices A (odd $y$) and B (even $y$). 
The hopping between these two sublattices lead to 
the coupled Harper equations,
\begin{eqnarray}
-t_1 \phi(y_A) -t_2 \phi(y_A+2) &=& E \phi(y_A+1),
\\
-t_2 \phi(y_A-1) -t_1 \phi(y_A+1) &=& E \phi(y_A),
\end{eqnarray}
where $y_A=1,3,5,...,L_y-1$ are the lattice points of sublattice A.
The open boundary condition in the $y$-direction requires
$\phi(L_y+1)=0$ and $\phi(0)=0$.
From the coupled Harper equations, the wavefunctions of the eigenstates 
and the corresponding energy spectrum can be obtained in an analytical form. 
For the chains with $t_1>t_2$ 
($|k_x| < 2 \pi / 3$), 
the wavefunction is given by 
\begin{equation}
\Psi(k_x; y_A) 
\equiv \left(\begin{array}{c}
\phi_{p_y}(y_A)
\\
\phi_{p_y}(y_A+1)
\end{array}
\right)
= \left(\begin{array}{c}
\pm \sin[p_y y_A + \varphi(p_y)]
\\
\sin[p_y(y_A+1)]
\end{array}
\right),
\label{ExtendedState}
\end{equation}
where the extra phase $\varphi$, chosen to be in the range 
$0 \le \varphi(p_y) < \pi/2$, depends on the magnitude of momentum $p_y$,
\begin{equation}
\varphi(p_y) = 
\tan^{-1} \left[ \frac{t_1-t_2}{t_1+t_2} \tan p_y \right].
\end{equation}
Finiteness of $L_y$ causes the quantization of $p_y$ 
which satisfies the following constraint
\begin{equation}
(L_y + 1) p_y + \varphi(p_y)= m \pi,
\label{ExtendMomentum}
\end{equation}
where $m$ is an integer. 
The energy spectrum for these states is given by 
\begin{equation}
E(k_x, p_y) = \pm \sqrt{t_1^2+t_2^2 + 2 t_1 t_2 \cos (2 p_y)}. 
\label{eq:Extendenergy}
\end{equation}
Here we emphasize that the momentum in the $y$-direction, 
$k_y$, is no longer a good quantum number due to the open boundaries. 
However, the magnitude of the momentum $p_y$ 
remains a good one and would be quite helpful for identifying 
dominating interactions later. 
Since the wavefunction Eq. (\ref{ExtendedState}) is extended into 
the bulk sites, we call the states ``extended states".

For the chains with $t_1<t_2$ 
($|k_x| > 2 \pi / 3$), 
peculiar localized states show up. 
One finds that there are only $L_y - 2$ extended states 
in Eq. (\ref{ExtendedState}).
The missing two states are linear combinations of localized states 
near the edges with complex momentum $k_y = \pm \pi/2 \pm i \gamma$. 
The wavefunctions for these edge states are
\begin{equation}
\Psi(k_x; y_A) = e^{i\pi y_A/2}\left(\begin{array}{c}
\pm \sinh[\gamma (L_y+1-y_A)]
\\
\sinh[\gamma(y_A+1)]
\end{array}
\right).
\label{EdgeState}
\end{equation}
The imaginary part of the momentum $\gamma \ne 0$ is the inverse of 
localization length of the edge states and satisfies the constraint
\begin{equation}
(L_y+1) \gamma 
= \tanh^{-1} \left[ - \frac{t_1+t_2}{t_1-t_2} \tanh \gamma \right].
\label{EdgeMomentum}
\end{equation}
It is clear that the nonzero solution of $\gamma$ only exists when $t_1<t_2$. 
The energy spectrum for these edge states is given by 
\begin{equation}
E(k_x,\gamma) = \pm \sqrt{t_1^2+t_2^2-2 t_1 t_2 \cosh(2 \gamma)}.
\label{eq:Edgeenergy}
\end{equation}

\begin{figure}
\centering
\includegraphics[width=60mm]{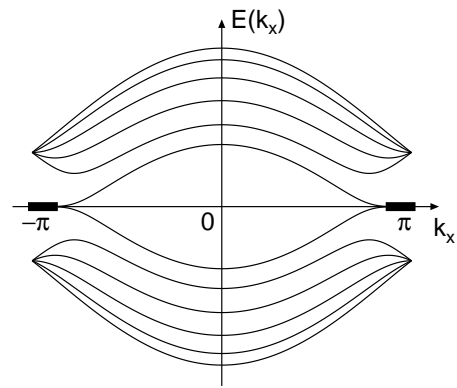}
\caption{Band structure of the zigzag NGR with $L_y = 12$.
The bold lines represent dispersions around the Fermi point, $k_x = \pi$.
}
\label{fig:NGRband}
\end{figure}

We thereby obtain the whole band structure of the tight-binding Hamiltonian,
Eqs. (\ref{eq:Extendenergy}) and (\ref{eq:Edgeenergy}).
The band structure for $L_y = 12$ is shown in Fig. \ref{fig:NGRband}, 
as an example.
It is worth noticing that the energy spectrum of the edge states 
is always lower than those of the extended states. 
It means that the edge states would dominate 
the low-energy physics in the weak-coupling limit 
while the extended states in the bulk do not play a crucial role 
because all of them are suppressed by a finite gap of order $t/L_y$.

\subsection{Renormalization group analysis}
Now, we are ready to treat the interacting Hamiltonian ${\cal H}_U$.
After the partial Fourier transformation, the interacting Hamiltonian 
is written as, 
\begin{eqnarray}
\tilde{\cal H}_U &=& 
\frac{U}{L_x} \sum_{\{k_{x_i}\}} \sum_{y=1}^{L_y}
\delta(k_{x_1}-k_{x_2}+k_{x_3}-k_{x_4})
\nonumber \\
& &
\times c^\dagger_\uparrow(k_{x_1}; y) c_\uparrow(k_{x_2}; y) 
       c^\dagger_\downarrow(k_{x_3}; y) c_\downarrow(k_{x_4}; y).
\label{eq:tHamU}
\end{eqnarray}
The delta function represents the momentum conservation in the $x$-direction.
The Hamiltonian can be rewritten in terms of the eigenstates of 
$\tilde{{\cal H}}_0$ by expanding the electron operator 
on the diagonal basis, 
\begin{equation}
c_\alpha(k_x; y) = \sum_p \phi_p (y) \psi_{p \alpha} (k_x).
\label{eq:def-of-field}
\end{equation}
where $\sum_p$ is taken for all the extended states $p = p_y$ and, 
if ever, the edge states $p = \pm \gamma$.
The resultant Hamiltonian is 
\begin{eqnarray}
\tilde{\cal H}_U &=& \frac{U}{L_x} \sum_{\{k_{x_i}\}} \sum_{\{p_i\}} 
\delta(k_{x_1}-k_{x_2}+k_{x_3}-k_{x_4}) 
\nonumber\\
&& \times \sum_{y} \big[ \phi^{*}_{p_1}(y) \phi^{}_{p_2}(y)  
\phi^{*}_{p_3}(y) \phi^{}_{p_4}(y)\big]
\nonumber\\
&& \times \psi^{\dag}_{p_1 \uparrow}(k_{x_1}) 
\psi^{}_{p_2 \uparrow}(k_{x_2}) 
\psi^{\dag}_{p_3 \downarrow}(k_{x_3})
\psi^{}_{p_4 \downarrow}(k_{x_4}).
\label{Vertices}
\end{eqnarray}
In general, $p_i$ are incommensurate and the dominant contribution 
comes from the pairwise equal $p_i$ vertices while other kinds of 
vertices only occur in a limited tiny phase space. 
This great reduction of dominant interactions enables us 
to classify them into three categories. 
(i) The interactions involving only the edge states: 
Since the normalized wavefunction of the edge states is order 
one near the edge and vanishingly small in the bulk, 
the wavefunction product of $\phi_{p_i}(y)$ in Eq. (\ref{Vertices}) 
after summation over $y$ is order one. 
Thus the edge-edge interaction remains ${\cal O}(U)$. 
(ii) The interactions involving a pair of the extended 
states and a pair of the edge states: 
Due to the spreading of wavefunction in the extended state, 
the summation of the wavefunction product in Eq. (\ref{Vertices}) is 
roughly $|\phi_{p_y}(1)|^2 \sim 1/L_y$.
The edge-bulk interaction is therefore only ${\cal O}(U/L_y)$ and 
becomes smaller as the width of the NGR grows. 
(iii) The interactions involving only the extended states:
From the similar argument to the case of the edge-bulk interaction,
the bulk-bulk interaction turns out to be ${\cal O}(U)$.
The effect of this interaction in the bulk system ($L_y \to \infty$)
has been studied in Refs. \onlinecite{Lin} and \onlinecite{BF} 
by the weak-coupling analysis.
It has turned out that the interaction opens a finite energy gap 
even if the noninteracting Hamiltonian $\tilde{\cal H}_0$ includes 
gapless chains with $k_x = \pm 2\pi/3$.

Let us move on to deriving the low-energy effective Hamiltonian 
in the continuum limit.
In the weak-coupling limit, the effective theory 
is dominated by the low-lying edge states near $k_x \approx \pi$ 
because all other bands in the bulk are gapped.
Thus, we may keep only the edge states and neglect all the gapped modes.
That is to say, only the kinetic energy of the edge modes 
and the edge-edge interaction term must be considered in the limit.
It is useful to construct the antibonding and bonding fields 
$\psi_{\pm \alpha}(x)$ from the field for the edge states defined 
in Eq. (\ref{eq:def-of-field}),
\begin{eqnarray}
\psi_{\pm \alpha}(x) &=& \int^{\Lambda}_{-\Lambda} \frac{dk_x}{2\pi} 
e^{ik_x x}\psi_{\pm \alpha}(k_x+\pi),
\\
\psi_{\pm \alpha}(k_x) &=& \frac{1}{\sqrt{2}}
[ \psi_{+\gamma \alpha}(k_x) \pm \psi_{-\gamma \alpha}(k_x) ],
\end{eqnarray}
where $\Lambda$ is a momentum cutoff.
Furthermore, we approximate the energy spectrum near $k_x \approx \pi$ 
to the lowest non-vanishing order $E_{\pm}(k) \simeq \pm c (k_x -\pi)^z$, 
where the $\pm$ signs stand for the antibonding and bonding edge modes 
and the dynamical exponent $z= L_y-1$. 
The kinetic energy can then be expressed in terms of the fields 
$\psi_{\pm \alpha}(x)$ in the continuum limit
\begin{equation}
\tilde{\cal H}_0 = \int dx \sum_{\mysw{P}=\pm} 
P \psi_{\mysw{P}\alpha}^{\dag}(x) 
(-i\partial_x)^z \psi_{\mysw{P}\alpha}^{}(x),
\label{Kinetic}
\end{equation}
where the constant $c$ is set to unity for notation simplicity. 

While it is rather straightforward to write down the kinetic terms 
in the form of Eq. (\ref{Kinetic}), expressing the edge-edge interaction 
in terms of the field operators is far from trivial. 
Our strategy to go from lattice to continuum limit is to express 
the lattice operator for electrons in the edge states, 
$\tilde{c}_\alpha(x,y)$, in terms of the field operator $\psi_{\pm \alpha}(k)$ 
and relate it to $\psi_{\pm \alpha}(x)$ by Fourier transformation. 
To achieve the goal, we need to express $\tilde{c}_\alpha(x,y)$ 
in the eigenbasis of $\tilde{\cal H}_0$,
\begin{equation}
\tilde{c}_\alpha(x,y) = \int \frac{dk_x}{2\pi} e^{ik_x x} 
\sum_{\mysw{P}=\pm} \phi^{*}_{\mysw{P}}(y) \psi_{\mysw{P} \alpha}(k).
\end{equation}
Here a difficulty comes from the entanglement of summation over 
$k_x$ and $P$. 
Since $\gamma$ depends on the longitudinal momentum $k_x$ as 
in Eq. (\ref{EdgeMomentum}), 
the wavefunction $\phi_\mysw{P}(y)$ also has an implicit dependence on $k_x$. 
As a result, the lattice operator $\tilde{c}_\alpha(x,y)$ is not simply 
related to the continuous field $\psi_{\pm \alpha}(k)$ 
by Fourier transformation which requires the change of order 
between the summation over $P$ and the integral over $k_x$.
However, since the edge states with $k_x \simeq \pi$ are localized near 
the edges, it is reasonable to adapt the ^^ ^^ sharp-edge" approximation 
which assumes that all the wavefunctions of the edge states 
take the simple form as for $k_x = \pi$, i.e., 
\begin{equation}
\phi_{\pm}(y) = \frac{1}{\sqrt{2}} (\delta_{y,1} \pm \delta_{y,L_y}).
\end{equation}
Within the sharp-edge approximation, the relation between 
the lattice and the field operators becomes very simple, 
\begin{eqnarray}
\tilde{c}_\alpha(x,1) &=&  
\frac{e^{i \pi x}}{\sqrt{2}}[\psi^{}_{+ \alpha}(x) + \psi^{}_{- \alpha}(x)],
\label{FieldDecomposition1}
\\
\tilde{c}_\alpha(x,L_y) &=& 
\frac{e^{i \pi x}}{\sqrt{2}}[\psi^{}_{+ \alpha}(x) - \psi^{}_{- \alpha}(x)],
\label{FieldDecomposition2}
\\
\tilde{c}_\alpha(x,y) &=& 
0~~~~~(2 \le y \le L_y-1).
\label{FieldDecomposition3}
\end{eqnarray}
By substituting Eqs. (\ref{FieldDecomposition1}) - 
(\ref{FieldDecomposition3}) into the edge-edge interacting Hamiltonian 
\begin{equation}
{\cal H}_{ee} = U \sum_{\bm r}
\tilde{c}^{\dag}_{\uparrow}(x,y) \tilde{c}^{}_{\uparrow}(x,y)
\tilde{c}^{\dag}_{\downarrow}(x,y) \tilde{c}^{}_{\downarrow}(x,y),
\nonumber
\end{equation}
we obtain its continuum counterpart,
\begin{eqnarray}
\tilde{\cal H}_{ee} &=& g_c (J_{+}^{2} + J_{-}^2) + g_{\rho} J_{+} J_{-} 
- g_{\sigma} \bm{J}_{+} \cdot \bm{J}_{-}
\nonumber\\
&+&  \frac{g_u}{2} (I_{+} I^{\dag}_{-}+I_{-} I^{\dag}_{+} ).
\label{CurrentInteraction}
\end{eqnarray}
Here the $SU(2)$ invariant currents are defined as 
\begin{eqnarray}
J_{\mysw{P}} &\equiv& \frac12 \psi^{\dag}_{\mysw{P}\alpha} 
                              \psi^{}_{\mysw{P}\alpha},
\\
\bm{J}_{\mysw{P}} &\equiv& \frac{1}{2} \psi^{\dag}_{\mysw{P}\alpha} 
\bm{\sigma}_{\alpha \beta} \psi^{}_{\mysw{P}\beta},
\\
I_{\mysw{P}} &\equiv& \frac12 \psi^{\dag}_{\mysw{P}\alpha} 
\epsilon_{\alpha \beta}\psi^{}_{\mysw{P}\beta},
\end{eqnarray}
where $\bm{\sigma}$ and $\epsilon_{\alpha \beta}$ denotes Pauli matrices 
and an antisymmetric tensor, respectively.
The bare values of these couplings generated by the on-site interaction $U$ 
are $g_{c}=g_{\rho}= g_{\sigma} = g_{u} = U$.
We note that Eq. (\ref{CurrentInteraction}) is the most general form of the 
interacting Hamiltonian which includes four-fermion interactions with 
preserving the $U(1) \times SU(2)$ symmetry and the momentum conservation.
In fact, one can use Eq. (\ref{CurrentInteraction}) to consider 
more complicated short-range interaction, 
which would generate a different set of bare couplings.

Finally, the low-energy field theory of the zigzag NGR in the continuum 
limit is described by Eqs. (\ref{Kinetic}) and (\ref{CurrentInteraction}). 
At the tree level, the scaling dimensions of the all four-fermion 
interactions are $d(g) = z-1$ by simple dimension counting. 
Since the dynamical exponent is greater than one for the NGR, 
all the couplings $g_{c}, g_{\rho}, g_{\sigma}$, and $g_{u}$ are 
relevant under the RG transformation. 
However, we note that relevance of these couplings does not necessarily 
warrant excitation gaps in the energy spectrum. 
To identify the ground state, it is necessary to apply 
another approach such as the bosonization technique.
We will discuss it in the next subsection.

\subsection{Mott insulating ground state}

In this subsection, we consider what we can derive from the low-energy 
effective Hamiltonian obtained in the previous subsection.
One clue is the fact that the effective Hamiltonian has a form 
strikingly similar to the usual effective theory for 1D systems 
except for the large dynamical exponent $z > 1$.
Hence, we discuss the analogy between the NGRs and a one-chain system 
to derive the ground-state properties of the NGR.
On the other hand, the spectrum of the low-lying edge states in the NGR 
can be viewed as a two-chain system, whose bonding and antibonding bands 
touch the Fermi energy at one Fermi point as discussed below.
Therefore, it is also useful to consider the NGR in the analogy with 
the two-chain system.
We will see that the both analyses give the same conclusion: 
the NGR has a spin-singlet ground state with finite charge and spin gaps. 

Let us start with the first approach, i.e., 
considering the NGR in the analogy to a one-chain system at half-filling.
In this analogy, the field $\psi_{\mysw{+} \alpha}$ 
($\psi_{\mysw{-} \alpha}$) is regarded as 
a right- (left-) moving field operator.
Since the dispersion of the fields $\psi_{\pm \alpha}(x)$ is not linear, the 
rigorous mapping from the fermionic theory to the bosonic one is not possible.
However, it is likely that the large dynamical exponent $z$ just serves to 
make all interactions relevant and we can still use the bosonization 
techniques to clarify physical properties of the ground states qualitatively. 
That is to say, we can treat the dynamical exponent $z=1+\epsilon$ 
as a perturbation and still use the conventional bosonization rules 
to study the physical properties of the NGRs. 
Under the assumption, we follow conventional bosonization steps 
and introduce two pairs of bosonic fields 
\begin{equation}
\psi_{\pm \alpha}(x) \simeq \sqrt{\rho}~ 
e^{i[\phi_{\alpha}(x)\pm \theta_{\alpha}(x)]/2},
\label{ContinuousField}
\end{equation}
where $\phi_\alpha(x)$ and $\theta_\alpha(x)$ are respectively the phase and 
displacement fluctuations, which obey the commutation relation, 
\begin{equation}
[ \phi_\alpha(x), \theta_{\alpha'}(x')] 
= -2\pi i [1 + {\rm sign}(x-x')] \delta_{\alpha \alpha'}.
\end{equation}
Furthermore, we switch to the charge and spin basis defined by 
\begin{equation}
(\phi_{\rho,\sigma}, \theta_{\rho,\sigma}) 
= \frac{1}{\sqrt{2}}[(\phi_{\uparrow}, \theta_{\uparrow}) 
\pm (\phi_{\downarrow}, \theta_{\downarrow})].
\end{equation}
The effective interaction which is relevant to gap formation in 
Eq. (\ref{CurrentInteraction}) can be written down in terms of 
these bosonic fields,
\begin{equation}
\tilde{\cal H}_{ee} = g_{\sigma} \cos(\sqrt{2} \theta_{\sigma}) 
- g_u \cos(\sqrt{2} \theta_{\rho}).
\end{equation}
The absence of their dual fields $\phi_{\rho}$ and $\phi_{\sigma}$ 
in the interactions comes from the charge and spin conservations.
For the on-site interaction, both $g_{\sigma}$ and $g_u$ are positive 
and grow under RG transformation. 
Consequently, the corresponding bosonic fields are pinned 
at specific values up to fluctuations, 
\begin{equation}
\sqrt{2} \theta_{\sigma} \simeq \pi, \qquad \sqrt{2} \theta_{\rho} \simeq 0.
\label{PinnedBoson}
\end{equation}
The signs of the couplings are of crucial importance here. 
Expanding the cosine term around the pinned values, 
the system acquires finite gaps in both charge and spin sectors. 
Therefore, the ground state is a Mott insulator 
with both charge and spin gaps.
The formation of the charge gap is not surprising because the electron 
density is commensurate with the underlying lattice structure. 
The finite spin gap indicates that the spin rotational symmetry is 
not broken spontaneously and the ground state is a spin singlet, 
as predicted by Lieb's theorem. 
The result differs from the mean-field theory predictions 
where the quantum fluctuations are ignored. 
We believe that the strong fluctuations in 1D systems would invalidate 
the mean-field approach and restore the symmetry which is broken 
at the mean-field level.

\begin{table}
\begin{tabular}{|c|c|c|c|}\hline\hline
phase&$(\sqrt{2}\theta_{\sigma}, \sqrt{2}\theta_{\rho})$& $C_s$ & $D$\\
\hline\hline
New&$(\pi,0)$&0&0\\
\hline
CDW&$(0,\pi)$&$\neq$ 0&0\\
\hline
Dimer&$(0,0)$&0&$\neq$ 0\\
\hline\hline
\end{tabular}
\caption{Comparison between different phases.}
\label{tab:pin}
\end{table}

It is worth noting that for the usual weakly-coupled 1D system 
with linear energy spectrum, the bosonic fields never get pinned at the values 
listed in Eq. (\ref{PinnedBoson}) because it is not stable in the RG flow. 
However, the new phase here is stable due to the large dynamical exponent 
which corresponds to the van Hove singularity in the density of states.
We summarize the pinned values for each phase in Table \ref{tab:pin}.
To differentiate this new phase from the familiar charge density wave 
or the dimer phase, one needs to compute the order parameters, i.e., 
charge density modulation $C_s$ and bond dimerization $D$ 
defined as
\begin{eqnarray}
C_{s} &\equiv& (-1)^x 
\langle a^{\dag}_{\alpha}(x) a^{}_{\alpha}(x) \rangle
\nonumber\\
&\simeq& \langle \psi^{\dag}_{\mysw{+}\alpha} \psi^{}_{\mysw{-}\alpha} \rangle 
+ \langle \psi^{\dag}_{\mysw{-}\alpha} 
\psi^{}_{\mysw{+}\alpha}  \rangle,
\\
D &\equiv& \frac{(-1)^x}{2} \langle a^{\dag}_{\alpha}(x) a^{}_{\alpha}(x+1) 
+ a^{\dag}_{\alpha}(x+1) a^{}_{\alpha}(x) \rangle
\nonumber\\
&\simeq& -i \langle \psi^{\dag}_{\mysw{+}\alpha} 
\psi^{}_{\mysw{-}\alpha}  \rangle 
+i \langle \psi^{\dag}_{\mysw{-}\alpha} 
\psi^{}_{\mysw{+}\alpha}  \rangle, 
\end{eqnarray}
where $a_\alpha (x)$ is the lattice operator in the one-chain system 
and related to the fields as 
$a_\alpha (x) \simeq e^{ik_\mysw{F} x} \psi_{\mysw{+} \alpha}(x)
+ e^{- ik_\mysw{F} x} \psi_{\mysw{-} \alpha}(x)$ with $k_\mysw{F} = \pi/2$.
Using the relations
\begin{eqnarray}
\psi^{\dag}_{\mysw{+}\uparrow}\psi^{}_{\mysw{-}\uparrow} &\simeq& 
i e^{-i\theta_{\rho}/\sqrt{2}} e^{-i\theta_{\sigma}/\sqrt{2}},
\\
\psi^{\dag}_{\mysw{+}\downarrow}\psi^{}_{\mysw{-}\downarrow} &\simeq& 
i e^{-i\theta_{\rho}/\sqrt{2}} e^{+i\theta_{\sigma}/\sqrt{2}},
\end{eqnarray}
and substituting the pinned values of the bosonic fields, 
one obtain the expectation values of the order parameters. 
The results are listed in Table \ref{tab:pin}. 
It is clear that there is no long-range order or symmetry breaking 
for the ground state of the zigzag NGRs.

\begin{figure}
\centering
\includegraphics[width=8cm]{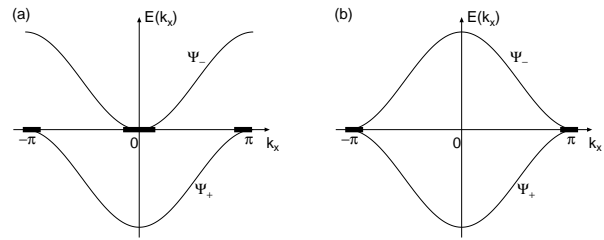}
\caption{Schematic picture of the band structure of 
(a) the two-chain system with $t_{\rm rung}/t_{\rm leg} \to 2$, 
and (b) the same system with the momentum shift $Q = \pi$ 
for the antibonding field.
The bold lines represent dispersions around Fermi points.
}
\label{fig:2band}
\end{figure}

Now we take another perspective and view the NGR 
as a system of two chains coupled via an interchain hopping.
Figure \ref{fig:2band} (a) shows the band structure of the two-chain system 
for the special point $t_{\rm rung}/t_{\rm leg} = 2$, 
where $t_{\rm rung}$ ($t_{\rm leg}$) 
is an amplitude of the interchain (intrachain) hopping. 
As the hoppings are tuned at the special value, the chemical potential 
cut through the bonding (antibonding) band at single Fermi point $k_x = \pi$ 
($k_x = 0$) with vanishing Fermi velocity $v_{\mysw{F}} \to 0$, 
rather than two pairs of Fermi points in general.
On the other hand, both the bonding and antibonding bands 
of the low-lying edge states in the NGRs 
have single Fermi point $k_x = \pi$ with $v_{\mysw{F}} \to 0$.
Hence, in the low-energy limit, the peculiar band structure in the NGRs 
can be viewed as a two-chain system with shifting the momentum of 
the antibonding field $k_x \to k_x + \pi$ 
[See Fig. \ref{fig:NGRband} and Fig. \ref{fig:2band} (b)]. 
One might worry that the dynamical exponent in the NGRs $z = L_y -1$ 
may be different from that in the two-chain system $z=2$.
However, the dynamical exponent $z$ in this limit only affects 
how the Fermi velocity approaches zero and should not 
give rise to qualitative difference.

In the weak-coupling limit, the lattice electron operator $a_{i \alpha}(x)$ 
in the $i$-th chain of the two-chain system can be described 
by two pairs of chiral fields
\begin{eqnarray}
a_{1\alpha}(x) &=& \frac{e^{i\pi x}}{\sqrt{2}} [\psi_{+\alpha}(x) 
+ e^{iQx}\psi_{-\alpha}(x)],
\label{TwoChain1}
\\
a_{2\alpha}(x) &=& \frac{e^{i\pi x}}{\sqrt{2}} [\psi_{+\alpha}(x) 
-e^{iQx} \psi_{-\alpha}(x)],
\label{TwoChain2}
\end{eqnarray}
where $Q=\pi$ is the momentum difference between Fermi points of 
bonding and antibonding bands in the two-chain system. 
The relation of lattice operators to the field operators 
for the two-chain system in Eqs. (\ref{TwoChain1}) and (\ref{TwoChain2}) 
are strikingly similar to those for the NGR 
in Eqs. (\ref{FieldDecomposition1}) and (\ref{FieldDecomposition2}), 
except the momentum shift $Q$. 
This enables us to translate the excitations and correlations 
in the two-chain system directly into those for the NGRs.

For on-site repulsive interaction, the ground state of the two-chain system 
is the Mott insulator with finite charge and spin gaps. 
In addition, there exists neither broken symmetry nor local order parameter. 
This Mott-insulating phase resembles the peculiar phase we discussed above 
by making analogy to the one-chain case. 
Thus, both approaches support that the ground state of the NGRs  
is a spin singlet with finite charge and spin gaps.

The two-chain analogy also gives us information about 
the spin-spin correlation.
It is known that the two-chain system has magnon excitations
with spin $S=1$ and momentum $P=(\pi, \pi)$. 
The momentum shift $Q$ in the $x-$direction shifts the low-lying magnon 
to $P=(0,\pi)$. 
This means that the spin-spin correlation functions in the NGRs 
are ferromagnetic within the same edge 
and antiferromagnetic between opposite edges.
This result is consistent with the fact that 
the sites in the same edge belong to the same sublattice 
while those in opposite edges belong to different sublattices.

\section{Numerical Results}

In this section, we present numerical results for the $\pi$-electron 
system (\ref{eq:Ham}) at half-filling. 
Hereafter, we set $t = 1$.
Using the DMRG method with improved algorithm,\cite{White1,White2} 
we have calculated various energy gaps and correlation functions.
The number of kept states is up to $m = 1200$ per block.
Numerical errors due to the truncation of the DMRG calculation 
are estimated from the difference between the data with different $m$'s.
We discuss the results for the zigzag NGRs 
and the zigzag CNTs in the following subsections.

\subsection{Zigzag nanographite ribbons}

\begin{figure}
\includegraphics[width=80mm]{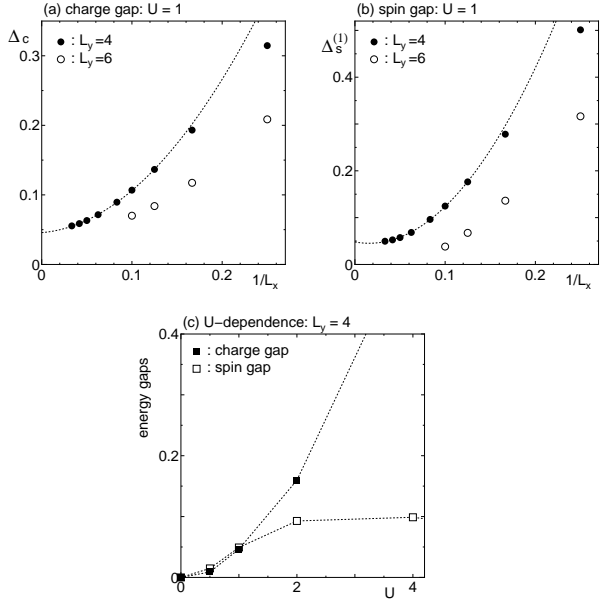}
\caption{
$L_x$-dependence of the energy gaps in the NGRs for $U = 1$; 
(a) the charge gap $\Delta_c$, (b) the spin gap $\Delta_{\rm s}^{(1)}$.
The solid and open circles represent the data for $L_y = 4$ and $L_y = 6$, 
respectively.
The numerical errors due to the DMRG truncation are smaller than symbols.
The gaps for $L_y = 4$ are extrapolated by fitting the data 
to a polynomial form, $\Delta(L_x) = \Delta(\infty) + a/L_x + b/L_x^2$ 
where $a$ and $b$ are fitting parameters.
(c) The $U$-dependence of the extrapolated gaps for $L_y = 4$.
} 
\label{fig:GRgap}
\end{figure}

In this subsection, we show results on the zigzag NGRs.
We first discuss the results for $L_y = 4$.
In all the cases calculated, we have found that 
the ground state belongs to 
the subspace of $M = 0$, where $M$ is the total magnetization in the system.
In Fig. \ref{fig:GRgap}, we show the data of the charge and spin gaps 
defined as
\begin{eqnarray}
\Delta_{\rm c} &=& \frac{1}{2}
             \left[E_0\left(\frac{N}{2}+1,\frac{N}{2}\right) 
                 + E_0\left(\frac{N}{2}-1,\frac{N}{2}\right) \right.
\nonumber \\
&&~~~~~~~ \left. - 2 E_0\left(\frac{N}{2},\frac{N}{2}\right) \right],
\label{eq:gapc} \\
\Delta_{\rm s}^{(M)} &=& E_0\left(\frac{N}{2}+M,\frac{N}{2}-M\right) 
             - E_0\left(\frac{N}{2},\frac{N}{2}\right),
\label{eq:gaps} 
\end{eqnarray}
where $E_0(N_\uparrow,N_\downarrow)$ is the lowest energy in the subspace 
of $N_\uparrow$ up-spin and $N_\downarrow$ down-spin electrons and 
$N$ is the total number of sites.
For $L_y = 4$, both the energy gaps are extrapolated to 
non-zero values at $L_x \to \infty$.
The $U$-dependence of the extrapolated gaps suggests that 
both of the gaps start to open from infinitesimal $U$, i.e., $U_c = 0$.
We thus conclude that for arbitrary $U > 0$ 
the ground state of the zigzag NGRs 
is a spin singlet with finite charge and spin gaps, 
being consistent with the weak-coupling analysis.

\begin{figure}
\includegraphics[width=80mm]{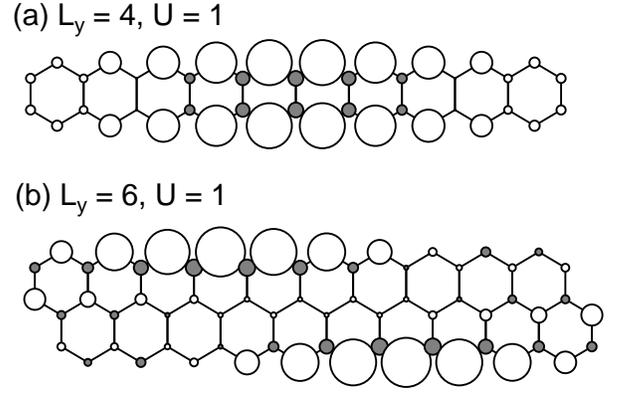}
\caption{
Distribution of local spin polarization $\langle S^z({\bm r}) \rangle_{M=1}$ 
in the NGRs with (a) $L_y = 4$ and (b) $L_y = 6$.
The Hubbard coupling and the total magnetization are 
$U = 1$ and $M = 1$.
Open and gray circles respectively represent the positive and negative 
values of the spin polarization while the areas of them are proportional 
to the magnitude.
} 
\label{fig:GRsz}
\end{figure}

To see how the spin polarizations are distributed in real space,
we have calculated the local spin polarization 
\begin{equation}
\langle S^z ({\bm r}) \rangle_M = 
\frac{1}{2} \langle n_\uparrow({\bm r}) - n_\downarrow({\bm r}) \rangle_M,
\end{equation}
where $\langle \cdots \rangle_M$ denotes the expectation value 
in the lowest-energy state in the subspace of $M$.
We have found that in the ground state 
the local spin polarization $\langle S^z ({\bm r}) \rangle_{M=0}$ 
is zero at all sites within the numerical error, 
supporting that the ground state is spin singlet.
There is no spontaneous spin polarization.
Figure \ref{fig:GRsz} (a) shows the distribution of 
the local spin polarization in the magnetized state of $M=1$ 
for $U = 1$ and $L_y = 4$.
As seen in the figure, the magnetization appears almost only 
at the edge sites.
This suggests that the electrons in the edge states are 
polarized by an external field more easily than those in the bulk.
We note that the localization character of the magnetization 
seems to persist even for a rather large coupling $U = 4$.
In other words, the edge states survive robustly 
against the electron-electron interactions 
even if $U$ becomes larger than the hopping amplitude $t$.

\begin{figure}
\includegraphics[width=60mm]{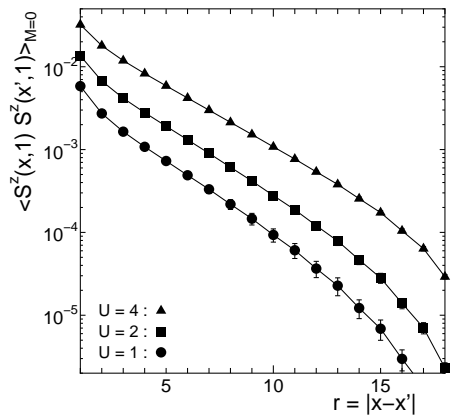}
\caption{
Ground-state spin-spin correlation functions in the NGRs with $L_y = 4$
and $L_x = 20$.
The circle, square, and triangle represent 
the data for $U = 1$, $2$, and $4$, respectively.
The error bars represent the numerical error due to the DMRG truncation.
} 
\label{fig:GRcor}
\end{figure}

In Fig.\ref{fig:GRcor}, we show the spin-spin correlation functions 
in the edge sites, 
\begin{equation}
\langle S^z(x,1) S^z(x',1) \rangle_{M=0}.
\end{equation}
To reduce boundary effects, the correlations are calculated 
between the sites of the same distance from the center of the NGR.
The correlations turn out to be always positive, 
suggesting a ferromagnetic coupling between the sites in the same edge.
We have also found that the spin correlation between the sites in opposite 
edges is always negative, i.e., antiferromagnetic.
These behaviors are consistent with the expectation 
from the two-chain analogy discussed above.
The correlations decay exponentially for all $U$'s reflecting the finite 
spin gap,\cite{boundary} and are enhanced as $U$ increases.

Finally, we discuss the dependence of low-energy properties 
of the zigzag NGRs on the width $L_y$.
Since the low-energy physics of the NGRs is dominated by 
the edge states which are strongly localized around the zigzag edges, 
it is likely that the singlet-triplet spin gap $\Delta_{\rm s}^{(1)}$, 
which corresponds to the energy scale required to magnetize the edge spins, 
comes from the antiferromagnetic effective coupling 
between the electrons in the edge states of opposite sides.
Thus, it is natural to expect that the spin gap decreases 
as the width $L_y$ increases.
To see this, we have performed calculations for the NGR 
with $L_y = 6$.\cite{GRbc} 
As can be seen in Fig. \ref{fig:GRsz} (b),
the spin polarizations in the magnetized state of $M = 1$ 
are found to be strongly localized at the zigzag edges, 
suggesting that the edge states exist also for $L_y = 6$.
Furthermore, we can clearly see in Fig. \ref{fig:GRgap} that 
the spin gap for the same $L_x$ 
decreases as the width of the ribbon $L_y$ increases, as expected.
Unfortunately, the system size calculated for $L_y = 6$ is not large enough 
to obtain the extrapolated gap $\Delta_{\rm s}^{(1)} (L_x \to \infty)$.
Calculations for larger systems are desirable 
to clarify the situation for the NGRs with larger $L_y$.

\subsection{Zigzag nanotubes with open edges}

For the zigzag CNTs with open ends, we can apply 
basically the same argument as that for the zigzag NGRs.
The difference between the two systems is merely the fact that 
in the CNTs the momentum along the $x$-direction is quantized as 
$k_x = 2\pi n/L_x$ due to the finite circumference $L_x$.
To see what happens, we have done calculations 
for the zigzag CNT with $L_x = 2$.
In this case, there is one edge state at each zigzag open end.
Although the width of the CNT is quite narrow, 
we will show that characteristic magnetic properties 
expected for general zigzag CNTs already appear.

\begin{figure}
\includegraphics[width=60mm]{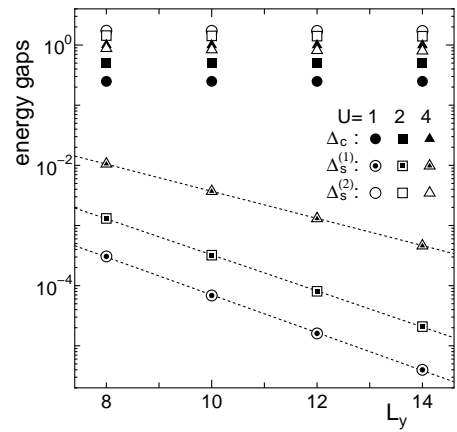}
\caption{
Charge gap $\Delta_{\rm c}$ and spin gaps 
$\Delta_{\rm s}^{(1)}$ and $\Delta_{\rm s}^{(2)}$ 
of the CNTs with $L_x = 2$
for $U = 1, 2$, and $4$ as functions of the length of CNTs $L_y$.
The numerical errors due to the DMRG truncation are smaller than symbols.
The dotted lines are guides for eye.
} 
\label{fig:NT2gap}
\end{figure}

We have found that the ground state always belongs to 
the subspace of $M = 0$ for all values of $U > 0$.
In Fig. \ref{fig:NT2gap}, we show the data of the charge gap $\Delta_{\rm c}$ 
and the spin gaps $\Delta_{\rm s}^{(1)}$ and $\Delta_{\rm s}^{(2)}$.
It can be seen in the figure that all the gaps take nonzero values 
in all the case of $U > 0$ and $L_y$ calculated.
We therefore conclude that the ground state of the zigzag CNTs 
with finite $L_y$ is a spin-singlet Mott insulator, same as the NGRs. 
An interesting observation here is that the spin gap $\Delta_{\rm s}^{(1)}$ 
decays exponentially as $L_y$ increases while 
$\Delta_{\rm s}^{(2)}$ converges to finite values at $L_y \to \infty$.
Hence, for large enough $L_y$, 
the magnetization $M = 1$ can be induced 
by applying an infinitesimal field 
whereas a finite field is needed to magnetize the system to $M \ge 2$.

\begin{figure}
\includegraphics[width=80mm]{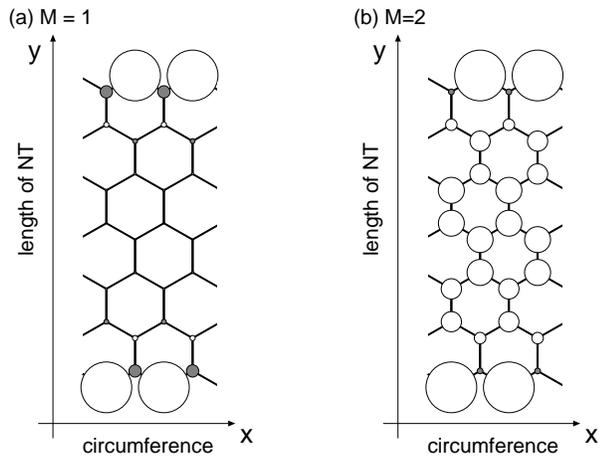}
\caption{
Distribution of local spin polarization in the CNTs with $L_x = 2$ 
for $U = 1$ and $L_y = 14$.
The total magnetization is (a) $M = 1$ and (b) $M = 2$.
Open and gray circles respectively represent the positive and negative 
values of the spin polarization while the areas of them are proportional 
to the magnitude.
} 
\label{fig:NT2sz}
\end{figure}

Next, we discuss how the spin polarizations are distributed in real space.
Same as the NGRs, the local spin polarization in the ground state 
is zero at all sites.
Figure \ref{fig:NT2sz} shows the distribution of 
the local spin polarization $\langle S^z({\bm r}) \rangle_M$ 
in magnetized states for $U = 1$.
The data clearly show that in the state of $M = 1$ 
the induced magnetization are strongly localized 
at the ends of the zigzag CNT.
This demonstrates the robustness 
of the localization character of the edge states 
against the coupling $U$.
In the state of $M = 2$, on the other hand, 
the distribution of $\langle S^z({\bm r}) \rangle_{M=2}$ 
seems to be a superposition of the edge magnetization 
and magnetization excited in the bulk sites.
We have found similar distributions of $\langle S^z({\bm r}) \rangle_M$ 
for $U = 2$ and $4$.

The results on the spin gaps and the local spin polarizations 
lead us to the conclusion that there are two different energy scales 
in magnetic excitations in zigzag CNTs: 
the energy to magnetize electrons in the edge states and 
the one to magnetize the bulk electrons.
As the length of the CNT increases, the former decays exponentially 
while the latter converges to a finite value, 
the spin gap of the bulk system.
This suggests that one can regard the edge states 
and the bulk electrons as two almost independent objects; 
The edge states are essentially 
decoupled from the bulk singlet state 
in spite of the presence of $U \simeq {\cal O}(t)$.
Very recently, it has been pointed out from a weak-coupling analysis 
that thicker CNTs including more than one edge states at each open ends 
also exhibit such a decoupling of the edge and bulk states.\cite{RyuH}
We therefore expect that the conclusion above is valid for 
general CNTs with zigzag open edges.

\subsection{Picture of the low-energy states}

From the above results, 
we can deduce a schematic picture to represent the low-energy physics 
of the zigzag NGRs and CNTs.
In this picture, the $\pi$-electron system consists of two parts: 
the electrons in the bulk forming a spin-singlet state 
and those in the edge states at each zigzag edge.
The electrons in the same edge are correlated ferromagnetically 
with each other to compose a large effective spin.
The effective spins interact via an effective antiferromagnetic coupling 
across the bulk singlet state.
Hence, the ground state of the system is a spin singlet in total. 
The effective coupling becomes smaller as the distance between 
the effective spins becomes larger, 
and finally, when the distance becomes large enough, 
the effective spins can flip freely, giving a paramagnetic response.
The bulk electrons remain forming a spin singlet unless a field 
corresponding to the bulk spin gap is applied.
We note that the picture above is basically consistent with 
the one discussed in Refs. \onlinecite{MFT2} and \onlinecite{PPP}.

\section{Discussion}

In this paper, we have studied low-energy properties of 
nanographite systems with zigzag edges in the presence of 
on-site Hubbard interactions $U$ using 
the weak-coupling RG and the DMRG method.
We have analyzed the Hubbard model on the zigzag NGRs 
and the zigzag CNTs.
We find that in both systems the ground state for $U > 0$ 
is a spin-singlet Mott insulator
with finite charge and spin gaps.
It is also found that the localization property of the edge state 
persists even for a rather large value of $U$, 
resulting in the effective spins localized around the zigzag edges.

Finally we wish to touch upon further extensions of this study.
One issue to be studied is effects of further hoppings and 
long-range electron-electron interactions.
As for magnetism, these long-range terms violate the assumptions 
of Lieb's theorem and open the possibility of exotic magnetic states.
However, it is also expected that the spin gap observed in the present study 
tends to stabilize the spin-singlet ground state against perturbations.
Very recently, it has been pointed out 
by the density-functional theory\cite{DFT2} 
that zigzag CNTs with a finite length can exhibit 
a high-spin ground state depending on the circumference $L_x$, 
although the method may underestimate effects of strong quantum fluctuation 
in the nanographite systems.
To clarify the situation, analyses with controlled approximations are needed.
Another issue is stacking effects.
Nanographite materials often take a stacking form, 
{\it e.g.}, nanographite particles composed by a few graphene sheets 
and multiwall CNTs.
Hence, it is important for applications 
to study effects of interlayer interactions.
For nanographite particles, it has been shown that 
their magnetic properties change drastically 
depending on types of stacking geometry.\cite{Hari}
In multiwall CNTs, the difference of radius and chirality 
between inner and outer layer is expected to affect not only the edge states 
but also the bulk properties in a complicated way.
Further extensive studies are desirable to tackle the issue.

\acknowledgements
We are grateful to K. Wakabayashi and A. Tanaka 
for fruitful discussions.
TH thanks S. Okada for sending a preprint before publication. 
TH and XH are supported by the Ministry of Education, Culture, 
Sports, Science and Technology, Japan, under the Priority Grant 
No. 14038240.
HHL and CYM appreciate the financial support from National Science Council 
in Taiwan through grant Nos. 91-2112-M007-040 (HHL), 91-2120-M-007-001 (HHL) 
and 91-2112-M007-049 (CYM).


\begin{thebibliography}{99}
\bibitem{C60}
H. Kroto {\it et al.}, Nature {\bf 318}, 162 (1985).

\bibitem{Iij}
S. Iijima, Nature {\bf 354}, 56 (1991).

\bibitem{review}
For a review, see R. Saito, G. Dresselhaus and M.S. Dresselhaus, 
{\it Physical Properties of Carbon Nanotubes} 
(Imperial College Press, 1998).

\bibitem{MFT1}
M. Fujita, K. Wakabayashi, K. Nakada and K. Kusakabe,
J. Phys. Soc. Jpn. {\bf 65}, 1920 (1996).

\bibitem{edge1}
K. Nakada, M. Fujita, G. Dresselhaus and M.S. Dresselhaus, 
Phys. Rev. B {\bf 54}, 17954 (1996).

\bibitem{edge2}
K. Wakabayashi, M. Fujita, H. Ajiki and M. Sigrist,
Phys. Rev. B {\bf 59}, 8271 (1999).

\bibitem{ACF1}
Y. Shibayama, H. Sato, T. Enoki and M. Endo, 
Phys. Rev. Lett. {\bf 84}, 1744 (2000).

\bibitem{ACF2}
Y. Shibayama, H. Sato, T. Enoki, X-X. Bi, M.S. Dresselhaus and M. Endo, 
J. Phys. soc. Jpn. {\bf 69}, 754 (2000).

\bibitem{el-ph}
M. Fujita, M. Igami and K. Nakada, J. Phys. Soc. Jpn. {\bf 66}, 1864 (1997).

\bibitem{MFT2}
K. Wakabayashi, M. Sigrist and M. Fujita, 
J. Phys. Soc. Jpn. {\bf 67}, 2089 (1998).

\bibitem{DFT1}
S. Okada and A. Oshiyama, Phys. Rev. Lett. {\bf 87}, 146803 (2001).

\bibitem{DFT2}
S. Okada and A. Oshiyama, unpublished.

\bibitem{Lieb}
E.H. Lieb, Phys. Rev. Lett., {\bf 62}, 1201 (1989).

\bibitem{White1}
S.R. White, Phys. Rev. Lett. {\bf 69}, 2863 (1992); 
Phys. Rev. B {\bf 48}, 10345 (1993).

\bibitem{Lin}
H.-H. Lin, Phys. Rev. B {\bf 58}, 4963 (1998).

\bibitem{BF}
L. Balents and M.P.A. Fisher, Phys. Rev. B {\bf 55}, R11973 (1997).


\bibitem{White2}
S. R. White, Phys. Rev. Lett. {\bf 77}, 3633 (1996).

\bibitem{Kennedy}
T. Kennedy, J. Phys. Condens. Matter {\bf 2}, 5737 (1990).

\bibitem{boundary}
We note that the suppression of the correlation functions 
from the linear behavior seen for large $r \sim L_x$ 
is due to the presence of the open boundaries and is not
essential.

\bibitem{GRbc}
For the NGR with $L_y = 6$, we remove the  boundary sites 
$(x,y) = (1, L_y)$ and $(L_x, 1)$, which are connected to other site 
by only one bond, to avoid undesirable boundary effects.
The shape of the system calculated is shown in Fig. \ref{fig:GRsz} (b).

\bibitem{RyuH}
S. Ryu and Y. Hatsugai, cond-mat/0211008.

\bibitem{PPP}
K. Nakada, M. Igami and M. Fujita, J. Phys. Soc. Jpn. {\bf 67}, 2388 (1998).

\bibitem{Hari}
K. Harigaya, Chem. Phys. Lett. {\bf 340}, 123 (2001).





\end{thebibliography}
\end{document}